\documentclass[12pt]{article}
\usepackage{amsmath,amsthm,amssymb}
\usepackage{german}
\theoremstyle{plain} 
\newtheorem{thm}{Theorem}
\newtheorem{cor}[thm]{Corollary}
\newtheorem{lem}[thm]{Lemma}
\newtheorem{prop}[thm]{Proposition}

\theoremstyle{definition}

\newcommand{\ncr}[2]{\mbox{$\left(\begin{array}{c}#1\\#2\end{array}\right)$}}

\title{
Invertibility of the TKF model of sequence evolution
\footnote{Published in  Mathematical Biosciences 200, no. 1 (2006) 58-75.}
}
\author{Bhalchandra D. Thatte \\
\small Allan Wilson Centre for Molecular Ecology and Evolution \\
\small Massey University, Palmerston North, New Zealand \\
\small \texttt{bdthatte@gmail.com} \\
}
\date{}

\begin{document}
\maketitle

\begin{abstract}
We consider character sequences evolving on a phylogenetic
tree under the TKF91 model. We show that as the sequence lengths
tend to infinity the the topology of the phylogenetic tree
and the edge lengths are determined by any one of
(a) the alignment of sequences (b) the collection of sequence lengths. 
We also show that the probability of any 
homology structure on a collection of sequences related by a TKF91
process on a tree is independent of the root location. 

\end{abstract}

\section {Introduction}
\label{sec:intro}

One of the important mathematical problems in phylogenetics
is formulated as follows.
Suppose a collection of sequences (of nucleotides
or amino acids) has evolved from a common ancestral
sequence under a stochastic process of evolution on a
phylogenetic tree. The process of evolution allows
substitution of characters, insertion or deletions
of segments of the sequence, and possibly other
types of mutations. A natural question is: given
current sequences (that is, sequences at the
leaves of the tree), can we reconstruct the
evolutionary relationship between them unambiguously?
Here evolutionary relationship means the topology of the underlying
phylogenetic tree and edge lengths, as well as
other parameters that define the stochastic process,
and possibly more information such as the homology
relationships between individual characters. One hopes
that given long enough sequences, the evolutionary relationship
between sequences is obtained accurately.

There are several reasons for studying such models.
Such models are a useful compromise between biological
reality and computational tractability.
They allow the use of maximum likelihood
methods for estimating evolutionary relationships,
and maximum likelihood methods can be proved to be
consistent for many such models.

Several earlier studies focused on pure substitution models only,
that is, models that didn't consider insertion-deletion
events. In one of the earlier papers by Felsenstein \cite{fel_1981},
an algorithm was proposed to compute the likelihood of
a tree (given sequences) for a class of reversible substitution
models. Many {\em invertibility} (unique identifiability of the
underlying tree) results have already been known for
classes of Markov substitution models. For example, in
\cite{ch_1991} and \cite{shp_1998} it was shown that
the tree topology and edge lengths are identifiable from
the joint distribution of character states at pairs of terminal
nodes under some mild conditions on the Markov transition matrices
on the edges of the tree. This result was proved using
the following result due to Hakimi and Patrinos \cite{hp_1972}
about {\em additive functions} on trees.
\begin{lem}
\label{additive}
Let $T$ be a tree on the vertex set $V$. Let $f$ be a
non-zero real valued function defined on the set of subsets of
$V$ of cardinality 2, satisfying the additivity condition
\begin{displaymath}
f(\{x,y\}) = \sum_{i=0}^{r-1} f(\{x_i,x_{i+1}\})
\end{displaymath}
where $x_0,x_1,\ldots , x_r$ is the unique path in $T$ connecting
$x=x_0$ and $y=x_r$. Then the value of $f$ on all pairs of
leaf nodes of $T$ determines uniquely the tree $T$ and
the function $f$.
\end{lem}
A restricted version of the above result for positive weights
was proved by Zaretskii \cite{zaretskii_1965} and
Buneman \cite{buneman_1971}.

General identifiability results in case of four-state
characters, which are of particular interest for
the phylogenetic analysis of DNA sequences, were
proved using a mathematical technique known as
{\em Hadamard Conjugation}, for example,
see \cite{shp_1998, hendy_1989, hp_1989, hp_1993}.
However, these results are based on certain
symmetry assumptions on transition matrices.
Strong results for general 12 parameter 4x4
Markov transition matrices have been proved by
Chang \cite{chang_1996}. He has proved that
full reconstruction (under some mild restrictions on
the Markov models) is possible by looking
at character distributions at triples of leaf
nodes.

The above mentioned results do not consider insertion-deletion
events. In 1991 and 1992, Thorne, Kishino and Felsenstein
\cite{tkf_1991, tkf_1992} introduced models that allowed
insertion-deletion events in addition to character substitution
events. We refer to the models by TKF91 and TKF92, respectively.
The models consider sequence evolution as
a birth-death process. In the TKF91 model, a birth event
inserts a single character in the sequence, and a death event
deletes a single character from the sequence. In the
TKF92 model, a birth event inserts  a segment
of characters in the sequence, and a death event deletes
a segment of characters from the sequence.
They proposed dynamic programming alignment algorithms
for a pair of sequences.
Recently the models have been extensively studied,
and many intricate algorithms for sequence alignment and
phylogenetic tree reconstruction have been proposed.
Although the algorithms based on the TKF framework are
difficult to describe and to implement, they have
many advantages, see for example, \cite{sh_2001, lmdjh_2003,
hjp_2003}. It is therefore important to analyse the TKF models
mathematically in the same way the symmetric and asymmetric
substitution models were analysed by Hendy, Penny, Steel and
Chang. In particular, it is worthwhile to know if these models are
invertible and if the algorithms based on them are
consistent.

In this paper, we prove two invertibility
results for the TKF91 model. We prove that an
alignment of sufficiently long multiple sequences,
(to be precise a homology structure on them,)
contains enough statistical information to estimate
the parameters of the birth-death process as well as
the tree topology and edge lengths.
We also show how to estimate the topology and model
parameters from only the sequence lengths provided the
sequences are sufficiently long. 
In the appendix, we prove that the probability of a general
homology structure on a collection of sequences does not
depend on the root location. This result is of theoretical
as well as practical importance. It has been used (without proof) 
to test implementations of algorithms in \cite{lmdjh_2003}.
Thus this paper gives a rigorous mathematical analysis of
the TKF91 model rather than proposing new algorithms.

The paper is structured as follows. In Section ~\ref{tkf91},
we give an overview of the TKF91 model, and demonstrate the
main idea of this paper for a pair of sequences.
In Section ~\ref{multiple}, we show the invertibility
based on the homology structure, and in Section~\ref{lengths}
we prove the invertibility based on sequence lengths.
In the last section we show the independence of the root
location and the likelihood of the homology structure,
and also discuss future directions.

\section{The TKF91 model}
\label{tkf91}
Let a DNA or an amino acid sequence be 
represented by an alternating string of links and
characters. The characters are denoted by \#
throughout, since the actual base or an amino acid
is not important in any discussion in this paper.
At the left end of the string is an immortal link.
All links undergo a birth process at a rate $\lambda$.
When a birth event occurs, a new character is placed
to the right of the link, and a new link is placed
to the right of the new character. All characters undergo
a death process at a rate $\mu $. When a character dies,
the character and the link to its right are deleted
from the string. The new character born in a birth
event is chosen from a distribution. While a character
is alive, it undergoes a substitution
process at a rate $s$, and the substituted character
is chosen from a distribution. The substitution model
and the model for the choice of the new character in a birth
event are not relevant here, since the homology structure
(information as to which characters are homologous
to which others) is what interests us in this paper.
Suppose that a sequence $A$ evolves to a sequence $B$
in time $t$. Since all characters at all times are assumed to be
evolving independently of all other characters,
we look at the evolution of a single character in $A$.
We describe the model by three coupled differential
equations and their solutions. We refer the reader to 
\cite{tkf_1991} for details.

Let $p_n^H(t)$ be the probability that a character survives
for time $t$, and at time $t$, it has $n$ descendents
(counting the survived character as one of the descendents).
Here superscript $H$ stands for ``homologous''.
By definition, $p_n^H(t) = 0$ for $n \leq 0$.
The differential equation for $p_n^H(t)$ is
\begin{equation}
\frac{dp_n^H}{dt} = \lambda (n-1)p_{n-1}^H + \mu n p_{n+1}^H
- (\lambda + \mu )n p_n^H
\end{equation}
with the initial conditions
\begin{eqnarray}
p_1^H(t=0) &=& 1 \nonumber \\ 
 p_n^H(t=0) &=& 0 \,\, \text{for}\,\, n > 1
\end{eqnarray}

Let $p_n^N(t)$ be the probability that the character in $A$
dies before time $t$, but leaves $n$ descendents in $B$.
Here superscript $N$ stands for ``non-homologous''.
By convention, $p_n^N(t) = 0$ for $ n < 0$. 
The differential equation for $p_n^N(t)$ is
\begin{equation}
\frac{dp_n^N}{dt} = \lambda (n-1) p_{n-1}^N + \mu (n+1) p_{n+1}^N
+ \mu p_{n+1}^H - (\lambda + \mu)np_n^N
\end{equation}
with the initial condition
\begin{equation}
p_n^N(t = 0) = 0 \,\,\text{for all}\,\,n
\end{equation}

Let $p_n^I(t)$ be the probability that the immortal link has
$n$ descendents at time $t$.
Here superscript $I$ stands for ``immortal''.
By convention, $p_n^I(t) = 0$ for $ n < 0$. 
The differential equation for $p_n^I(t)$ is
\begin{equation}
\frac{dp_n^I}{dt} = \lambda n p_{n-1}^I + \mu (n+1) p_{n+1}^I
- \lambda (n+1) p_{n}^I - \mu n p_n^I \,\, \text{for} \,\, n > 0
\end{equation}
with the initial condition
\begin{equation}
p_n^I(t = 0) = 0 \,\,\text{for all}\,\,n > 0
\end{equation}

\begin{lem}
\label{lem-sol}
Solutions to the above differential equations are given by
\begin{eqnarray}
\label{eq-sol}
p_n^H(t) &=& e^{-\mu t}( 1-\lambda \beta(t) ) (\lambda \beta(t) )^{n-1}
\,\,\text{for}\,\, n > 0
\nonumber \\
p_n^N &=& \mu \beta(t) \,\,\text{for}\,\, n = 0 \nonumber \\
&=& (1 - e^{-\mu t} - \mu \beta(t))( 1-\lambda \beta(t) ) 
(\lambda \beta(t) )^{n-1} \,\,\text{for}\,\, n > 0 \nonumber\\
p_n^I & = & ( 1-\lambda \beta(t) ) (\lambda \beta(t) )^{n}
\,\,for \,\,n \geq 0
\end{eqnarray}
where
\begin{equation}
\beta(t) = \frac{1 - e^{(\lambda - \mu)t}}
{\mu - \lambda e^{(\lambda - \mu )t}}
\end{equation}
\end{lem}
The main idea of this paper is the following observation
about the solutions.
\begin{cor}
\label{cor-pair-dist}
\label{param}
Let $A$ and $B$ be two sufficiently long sequences related
by the TKF91 model, $A$ being the ancestor of $B$.
Then model parameters (scaled time and the ratio of birth and death
rates) are uniquely determined by the true alignment of $A$ and $B$.
\end{cor}
\begin{proof}
Observe that $e^{-\mu t}$ is simply the probability that
a character in $A$ survives in $B$, $e^{-\mu t}\lambda \beta(t)$
is $\sum_{n \geq 2}p_n^H$, and $1-e^{-\mu t} - \mu \beta(t)$
is $\sum_{n \geq 1}p_n^N$. This allows us to accurately estimate
$\mu t$, $\lambda \beta(t)$ and $\mu \beta(t)$, and therefore
$\lambda / \mu$, from an alignment of sufficiently long
sequences.
\end{proof}

In the next section, this idea, together with Lemma~\ref{additive},
is applied to an alignment of several sequences.
One has to show that probabilities of certain patterns in the
alignment do not depend on the position of the root on the tree. 
Since not all patterns are required for estimating the pairwise
distances between leaves, the independence of the probability
of a pattern and the root position for general patterns is
presented in the Appendix. 

\section{Multiple sequences}
\label{multiple}
The main result of this section is the following.
\begin{thm}
\label{invert}
A multiple sequence alignment of sufficiently
long sequences that have evolved under a TKF91
process on a phylogenetic tree uniquely determines the
phylogenetic tree and the (scaled) edge lengths.
\end{thm}

Before presenting the full proof, it will be useful
to look at the main ideas behind the proof. 

\begin{enumerate}
\item To be able to apply Lemma ~\ref{additive} to construct the tree,
we would like to compute the distance between each pair of vertices.
Here the distance between a pair of vertices is the (scaled) divergence time
between the corresponding taxa. 

\item The distance between a pair of vertices is estimated
by the method of Corollary~\ref{cor-pair-dist}. This is
done by restricting the multiple alignment to the two sequences
under consideration and then equating the observed pattern
frequencies with those expected from the solutions of the
TKF model.

\item Before we can equate the observed frequencies of patterns
and the expected frequencies, we have to solve
one important problem: in the discussion of the TKF91 model,
we assumed that the sequence $A$ is the ancestor of the
sequence $B$. But if the sequences $A$ and $B$ evolve
from a common ancestor $C$ that is not $A$ or $B$, do we
expect the same pattern frequencies?
The bulk of the proof is devoted to answering this question by
showing that the pattern probabilities do not depend on the location
of $C$.
\end{enumerate}

Let $S_i; \, 1 \leq i \leq k$ be the sequences at the leaves
of a phylogenetic tree $T$ with root vertex $v$. The sequences
are related by a TKF91 process on the tree. The vertices of
the tree are labelled $v_i$ such that the first $k$ of them
are leaf vertices respectively associated with the sequences $S_i$.
Each vertex $v_i$ has associated with it a time parameter $t_i$,
which is the duration for which the TKF91 process operates on
the edge leading to $v_i$ from its immediate ancestor $a(v_i)$.

Consider any two sequences, say $S_1$ and $S_2$,
and let $v_i$ be their most recent common ancestor.
That is, the ancestral sequence $S$ at the root $v$
evolves into a sequence $S_i$ along the path from $v$
to $v_i$, and then two copies of $S_i$ evolve independently
to sequences $S_1$ and $S_2$, respectively, along paths
from $v_i$ to $v_1$ and $v_i$ to $v_2$.
It is assumed that the unknown ancestral sequence at the root
has already evolved for an infinite duration. It implies
that the sequence $S_i$ has evolved for an infinite
duration. Therefore, the homology relationship between $S_1$ and $S_2$
does not depend on the location of $v$, but perhaps
depends on the location of $v_i$ on the path between $v_1$
and $v_2$. Therefore, the analysis of the homology structure
of $S_1$ and $S_2$ may be performed by assuming that
$v_i$ is the same as the root $v$.
In other words, we will first analyse only two sequences
evolving from their most recent common ancestor, which is
assumed to have evolved for infinite time.

\subsection*{The restricted alignment and its blocks}
Consider the alignment of $S_1$ and $S_2$ that is
obtained by discarding all other rows in the multiple
alignment (sequences $S_3$ to $S_k$), and then discarding
the columns containing only gaps in the first two rows (rows for
$S_1$ and $S_2$). The most recent common ancestor of $v_1$ and $v_2$
is $v$. Since we have discarded other sequences,
and pruned the tree, we can assume that $v$ and $v_1$
are separated by time $t_1$, and  $v$ and $v_2$ are
separated by time $t_2$. We are interested in showing that the
probabilities of the patterns of interest are functions of $t_1 + t_2$,
that is, they do not depend individually on
$t_1$ or $t_2$.

An alignment of two sequences can be thought of as a
sequence of {\em blocks} where a block consists of
a pair of homologous characters followed by
non-homologous characters, and terminated by a pair
of homologous characters of the next block.
For example, a possible alignment of $S_1$ and $S_2$,
ignoring the beginning and end is
\begin{displaymath}
\begin{pmatrix}
S_1 \,\, \# & \# & \# & \# & \# & -  & \# & -  & \# & \# \\
S_2 \,\, \# & \# & -  & \# & -  & \# & \# & \# & -  & \#
\end{pmatrix}
\end{displaymath}
The alignment shows four types of blocks.
\begin{displaymath}
A \equiv
\begin{pmatrix}
\# & \# \\
\# & \#
\end{pmatrix}
\hspace{0.1in}
B\equiv
\begin{pmatrix}
\# & \# & \#\\
\# & -  & \#
\end{pmatrix}
\hspace{0.1in}
C\equiv
\begin{pmatrix}
\# & \# & - & \# \\
\# & - & \# &\#
\end{pmatrix}
\hspace{0.1in}
D\equiv
\begin{pmatrix}
\# & - & \# & \# \\
\# & \# & - &\#
\end{pmatrix}
\end{displaymath}

Homologous characters in a column imply that
there was a character in ancestral sequence $S$ that survived
in both $S_1$ and $S_2$.
We are interested in computing the probabilities
$p(A)$, $p(B)$, $p(C)$ and $p(D)$ of observing,
respectively, blocks of types $A$, $B$, $C$ and $D$.

\subsection*{The classes of events that result in the blocks of types 
$A$, $B$, $C$ and $D$.}

Computing the probabilities involves summing over
infinitely many possible events in the history
that result in a block of a certain type.
For example, a block of type $A$ is a result of
historical events of the type

\begin{displaymath}
\begin{pmatrix}
S_1 & \# & - & \# \\
S & \# & \#^i & \# \\
S_2 & \# & - & \#
\end{pmatrix}; i \geq 0
\end{displaymath}
Here we have $i$ characters in the ancestral sequence $S$
that died along the edges $v-v_1$ and $v-v_2$. Of course,
we do not know $S$. So looking at the alignment of $S_1$ and
$S_2$, we do not know how many characters died resulting
in a block of type $A$.
So the probability of observing $A$ would be the sum 
of probabilities over all non-negative values of $i$.

Similarly, there are two types of events that
result in a block of type $B$. They are
\begin{displaymath}
E_B^1 \equiv
\begin{pmatrix}
S_1 & \# & - & \# & - & \# \\
S & \# & \#^i & \# & \#^j & \# \\
S_2 & \# & - & - & - & \#
\end{pmatrix}
\,\,\text{and}\,\,
E_B^2 \equiv
\begin{pmatrix}
S_1 & \# & - & \# & - & \# \\
S & \# & \#^i & - & \#^j & \# \\
s2 & \# & - & - & - & \#
\end{pmatrix}
\end{displaymath}
where $i \geq 0, j \geq 0$.
Therefore, we write $p(B) = p(E_B^1) + p(E_B^2)$.
Here $E_B^1$ is the possibility that the middle
character in $S_1$ was homologous to a character in
$S$, but along the branch from $v$ to $v_2$, the
character in $S$ died. It is also possible
that the middle character in $S_1$ in block $B$ was created in
a birth event during the evolution from $S$ to $S_1$.
This is the possibility $E_B^2$.

The following four types of events result in a block of type $C$.
\begin{displaymath}
E_C^1 \equiv
\begin{pmatrix}
S_1 & \# & -    & \# & -    & -  &  -   &  \# \\
S & \# & \#^i & \# & \#^j & \# & \#^k &  \# \\
S_2 & \# & -    & -  & -    & \# &  -   &  \#
\end{pmatrix}; i \geq 0, j \geq 0, k\geq 0
\end{displaymath}
\begin{displaymath}
E_C^2 \equiv
\begin{pmatrix}
S_1 & \# & -    & \# & -    & -  &  -   &  \# \\
S & \# & \#^i & \# & \#^j & -  & \#^k &  \# \\
S_2 & \# & -    & -  & -    & \# &  -   &  \#
\end{pmatrix}; i \geq 0, j \geq 0, k\geq 0
\end{displaymath}
\begin{displaymath}
E_C^3 \equiv
\begin{pmatrix}
S_1 & \# & -    & \# & -    & -  &  -   &  \# \\
S & \# & \#^i & -  & \#^j & \# & \#^k &  \# \\
S_2 & \# & -    & -  & -    & \# &  -   &  \#
\end{pmatrix}; i \geq 0, j \geq 0, k\geq 0
\end{displaymath}
\begin{displaymath}
E_C^4 \equiv
\begin{pmatrix}
S_1 & \# & -    & \# & -    & -  &  -   &  \# \\
S & \# & \#^i & -  & \#^j & -  & \#^k &  \# \\
S_2 & \# & -    & -  & -    & \# &  -   &  \#
\end{pmatrix}; i \geq 0, j \geq 0, k\geq 0
\end{displaymath}

Observe that they are disjoint classes of events,
so we write 
\begin{displaymath}
p(C) = p(E_C^1) + p(E_C^2) + p(E_C^3) + p(E_C^4)
\end{displaymath}

Classes of events resulting in blocks of types $D$, 
are similar to those that produce the pattern
$C$. But there is no distinction between types $C$ and $D$
when $j=0$ in $E_C^4$. In the language of \cite{lmdjh_2003},
$C$ and $D$ have the same {\em homology structure}.
Therefore, we compute
\begin{displaymath}
p(C\vee D) = p(C) + p(D) - p(E_C^4; j = 0)
\end{displaymath}

The probability computations involve summations over
$i,j,k$ varying over all non-negative integers, but
the summations can be easily computed since they happen to be
geo\-metric series.

\subsection*{The Markov structure along the sequences}
How do we compute the probability of a
class of events contributing to a block?
The probability of an alignment of sequences on a tree is
the product of the probabilities of the pairwise
alignments on the edges of the tree. Also,
an alignment of two sequences has a Markov structure
along the sequences provided one of the sequences is the
ancestor of the other, see \cite{hjp_2003} for details.
We use this idea to compute probabilities of
classes of events using the Markov structure on
$v-v_1$ and $v-v_2$. 

The transition matrix for the Markov chain for
a two sequence alignment, assuming one of the
sequences to be the ancestor, is given by
\begin{eqnarray}
\label{markov}
\bordermatrix{ & H & D & I & end\cr \vspace*{2mm}
H   & (1- \lambda \beta)(\frac{\lambda}{\mu})e^{-\mu t} & 
(1- \lambda \beta)(\frac{\lambda}{\mu})(1-e^{-\mu t})   &
\lambda \beta &
(1- \lambda \beta)(1-  \frac{\lambda}{\mu}) \cr \vspace*{2mm}
D   & (1-\kappa)(\frac{\lambda}{\mu})e^{-\mu t} &
(1-\kappa)(\frac{\lambda}{\mu})(1-e^{-\mu t})   &
\kappa &
(1-\kappa )(1-  \frac{\lambda}{\mu}) \cr \vspace*{2mm}
I   &(1- \lambda \beta)(\frac{\lambda}{\mu})e^{-\mu t} &
(1- \lambda \beta)(\frac{\lambda}{\mu})(1-e^{-\mu t})  &
\lambda \beta &
(1- \lambda \beta)(1-  \frac{\lambda}{\mu}) \cr}
\end{eqnarray}
where $H$ ({\em homology}), $D$ ({\em deletion}) and
$I$ ({\em insertion}) denote the three states $\ncr{\#}{\#}$,
$\ncr{\#}{-}$ and $\ncr{-}{\#}$, respectively, and
\begin{displaymath}
\kappa(t) = \kappa = 1 - \frac{\mu \beta}{1 - e^{-\mu t}}
\end{displaymath}

Now on we denote $\beta(t_i)$ by $\beta_i$, $e^{-\mu t_i}$ by $X_i$,
$\kappa(t_i)$ by $\kappa_i$; for $i \in \{1,2\}$,
and $\beta(t) = \beta(t_1 + t_2)$ by $\beta$,
$e^{-\mu t}$ by $X= X_1X_2$, and $\kappa(t) = \kappa(t_1+t_2)$
by $\kappa$.

\subsection*{Computing $p(A)$}
Using the transition matrix in~(\ref{markov}), we
can write
\begin{eqnarray}
p(A) & = & (1-\lambda \beta_1)(1-\lambda \beta_2)
\left(\frac{\lambda}{\mu}\right)X_1X_2  \nonumber \\
& + & (1-\lambda \beta_1)(1-\lambda \beta_2)
\left(\frac{\lambda}{\mu}\right)(1-X_1)(1-X_2) \nonumber \\
& \times &
\sum_{i\geq 1}\left(\left(\frac{\lambda}{\mu}\right)(1-\kappa_1)(1-\kappa_2)
(1-X_1)(1-X_2)\right)^{i-1}   \nonumber \\
& \times & (1-\kappa_1)(1-\kappa_2)
\left(\frac{\lambda}{\mu}\right)X_1X_2   \nonumber \\
& = &
\frac{(1-\lambda \beta_1)(1-\lambda \beta_2)(\frac{\lambda}{\mu})e^{-\mu (t_1+t_2)}}{1 - (\frac{\lambda}{\mu})(1-\kappa_1)(1-\kappa_2)
(1-X_1)(1-X_2)}
\end{eqnarray}
where the first term corresponds to $i=0$.\\
One can verify that
\begin{eqnarray} 
\frac{(1-\lambda \beta_1)(1-\lambda \beta_2)}
{1 - (\frac{\lambda}{\mu})(1-\kappa_1)(1-\kappa_2)
(1-X_1)(1-X_2)} 
& = & \frac{(1-\lambda \beta_1)(1-\lambda \beta_2)}
{1- \lambda \mu \beta_1 \beta_2} \nonumber \\
& = & 1 - \lambda \beta
\end{eqnarray}
Therefore, 
\begin{equation}
\label{eq_a}
p(A) = (1 - \lambda \beta )\left(\frac{\lambda}{\mu}\right) e^{-\mu t}
\end{equation}

Thus, as required, $p(A)$ depends only on $t = t_1 + t_2$ but
not on $t_1$ and $t_2$ individually.

\subsection*{Computing $p(B)$}
We compute $p(E_B^1)$ as a product two factors $F_1$ and
$F_2$ which, respectively, correspond to the transitions shown below.

\begin{displaymath}
\begin{pmatrix}
1 & \# & - & \#  \\
v & \# & \#^i & \# \\
2 & \# & - & -
\end{pmatrix}; i \geq 0 \,\, \mbox{and} \,\,
\begin{pmatrix}
1 & \# & - & \#  \\
v & \# & \#^j & \# \\
2 & - & - & \#
\end{pmatrix}; j \geq 0
\end{displaymath}
Therefore,
\begin{eqnarray}
F_1 & = &
(1-\lambda \beta_1)(1-\lambda \beta_2)
\left(\frac{\lambda}{\mu}\right)X_1(1-X_2)  \nonumber \\
& + & (1-\lambda \beta_1)(1-\lambda \beta_2)
\left(\frac{\lambda}{\mu}\right)(1-X_1)(1-X_2) \nonumber \\
& \times & 
\sum_{i\geq 1} \left((1-\kappa_1)(1-\kappa_2)(\frac{\lambda}{\mu})
(1-X_1)(1-X_2)\right)^{i-1}  \nonumber \\
& \times & (1-\kappa_1)(1-\kappa_2)
\left(\frac{\lambda}{\mu}\right)X_1(1-X_2)  \nonumber \\
& = & (1 - \lambda \beta )
\left(\frac{\lambda}{\mu}\right) X_1(1-X_2)
\end{eqnarray}
and,
\begin{eqnarray}
F_2 = \frac{(1-\lambda \beta_1)(1-\kappa_2)
(\frac{\lambda}{\mu}) X_1X_2}
{1 -  \lambda \mu \beta_1 \beta_2}
\end{eqnarray}
Therefore,
\begin{eqnarray}
p(E_B^1) = \frac{(1 - \lambda \beta )(1-\lambda \beta_1)(1-\kappa_2)
(\frac{\lambda}{\mu})^2X_1^2X_2(1-X_2)}
{1 -  \lambda \mu \beta_1 \beta_2}
\end{eqnarray}
Next, the contribution from the class of events $E_B^2$ is
computed.
\begin{eqnarray}
p(E_B^2) & = & 
\left(\lambda \beta_1(1-\lambda \beta_2)
+ \frac{(1-\lambda \beta_1)(1-\lambda \beta_2)(\frac{\lambda}{\mu})(1-X_1)(1-X_2)\kappa_1(1-\kappa_2)}{1 -  \lambda \mu \beta_1 \beta_2} \right) \nonumber \\
&\times &\frac{(1-\lambda \beta_1)(\frac{\lambda}{\mu})X_1X_2}
{1 -  \lambda \mu \beta_1 \beta_2}
\end{eqnarray}
Adding $p(E_B^1)$ and $p(E_B^2)$, and after simplification,
we have
\begin{equation}
\label{eq_b}
p(B) = (1 - \lambda \beta )\lambda \beta
\left(\frac{\lambda}{\mu}\right)e^{-\mu t}
\end{equation}
Again, $p(B)$ depends only on $t = t_1 + t_2$.

\subsection*{Computing $p(C\vee D)$}
This computation is performed on similar lines as
the computation for $p(A)$ and $p(B)$ except that
the events in $E_C^4$ when $j=0$ also contribute to
$p(E_D^4)$. Therefore,
\begin{eqnarray}
\label{c_or_d}
p(C \vee D) & = & p(E_C^1)+ p(E_D^1) + p(E_C^2)+ p(E_D^2) + 
p(E_C^3)+p(E_D^3) \nonumber \\
&+& p(E_C^4)+p(E_D^4) - p(E_C^4; j = 0)
\end{eqnarray}
Skipping the details, we state below different terms
that contribute to $p(C\vee D)$.
\begin{eqnarray}
\label{eq_c1}
p(E_C^1) + p(E_D^1) & = &
2(1 - \lambda \beta )^2 \left(\frac{\lambda}{\mu}\right)e^{-\mu t}
\left(\frac{\lambda}{\mu}\right)e^{-\mu t}\frac{\lambda \mu \beta_1 \beta_2}
{1 -  \lambda \mu \beta_1 \beta_2}
\end{eqnarray}
\begin{eqnarray}
\label{eq_c2}
p(E_C^2)+p(E_D^2) = (1 - \lambda \beta )^2 \left(\frac{\lambda}{\mu}\right)
e^{-\mu t}\left(\frac{\lambda}{\mu}\right)
\left(\frac{X_1 (1-X_2)\kappa_2 + X_2(1-X_1)\kappa_1}
{1 - \lambda \mu \beta_1 \beta_2}\right)
\nonumber \\
\end{eqnarray}

\begin{eqnarray}
\label{eq_c3}
\lefteqn{p(E_C^3)+p(E_D^3)}\nonumber \\[5mm]
 & = &
(1 - \lambda \beta )^2 \left(\frac{\lambda}{\mu}\right)
e^{-\mu t}\left(\frac{\lambda}{\mu}\right)
\left((1-X_1)X_2\kappa_1+(1-X_2)X_1\kappa_2\right)
\left(\frac{\lambda \mu \beta_1 \beta_2}{1 - \lambda \mu \beta_1 \beta_2}
\right)  \nonumber \\[5mm]
& + & (1 - \lambda \beta ) \left(\frac{\lambda}{\mu}\right)
e^{-\mu t} \left(\frac{\lambda}{\mu}\right)
\frac{\lambda \mu \beta_1^2X_2(1 - \lambda \beta_2 )
+ \lambda \mu \beta_2^2X_1(1 - \lambda \beta_1 )
}{1 - \lambda \mu \beta_1 \beta_2}
\end{eqnarray}

\begin{eqnarray}
\label{eq_c4}
\lefteqn {p(E_C^4)+p(E_D^4) - p(E_C^4; j = 0) }\nonumber \\[5mm]
&=&(1 - \lambda \beta )^2 \left(\frac{\lambda}{\mu}\right)
e^{-\mu t}\left(\frac{\lambda}{\mu}\right)
(1-X_1)(1-X_2)\kappa_1 \kappa_2 \nonumber \\[5mm]
& + & (1 - \lambda \beta ) \left(\frac{\lambda}{\mu}\right)
e^{-\mu t} \left(\frac{\lambda}{\mu}\right)
\lambda \mu \beta_1 \beta_2 \nonumber \\[5mm]
& + & 2(1 - \lambda \beta )^2 \left(\frac{\lambda}{\mu}\right)
e^{-\mu t}\left(\frac{\lambda}{\mu}\right)
(1-X_1)(1-X_2)
\times \kappa_1 \kappa_2
\left(\frac{\lambda \mu \beta_1 \beta_2}{1 - \lambda \mu \beta_1 \beta_2}
\right) \nonumber \\[5mm]
& + & (1 - \lambda \beta )\left(\frac{\lambda}{\mu}\right)
e^{-\mu t}\left(\frac{\lambda}{\mu}\right)(1-X_1)(1-X_2) \nonumber \\[5mm]
&\times &\frac{\lambda \beta_1(1-\lambda \beta_2)(1-\kappa_1)\kappa_2
+ \lambda \beta_2(1-\lambda \beta_1)(1-\kappa_2)\kappa_1}
{1 - \lambda \mu \beta_1 \beta_2}
\end{eqnarray}

Adding Equations ~(\ref{eq_c1}), ~(\ref{eq_c2}), ~(\ref{eq_c3}),
and ~(\ref{eq_c4}), and simplifying, we get
\begin{eqnarray}
\label{eq_c_or_d}
p(C\vee D) & = &(1 - \lambda \beta )^2 \left(\frac{\lambda}{\mu}\right)
(1 - e^{-\mu t} - \mu \beta) \left(\frac{\lambda}{\mu}\right) e^{-\mu t}
\nonumber \\ [3mm]
& + &\lambda \beta (1 - \lambda \beta ) \left(\frac{\lambda}{\mu}\right)
\mu \beta \left(\frac{\lambda}{\mu}\right)e^{-\mu t} \nonumber \\[4mm]
& = & (1 - \lambda \beta )^2 
\left (\frac{\lambda}{\mu} - \frac{\lambda}{\mu}e^{-\mu t} - \lambda \beta
\right ) \left(\frac{\lambda}{\mu}\right)e^{-\mu t}
+ (\lambda \beta)^2 (1 - \lambda \beta) \left(\frac{\lambda}{\mu}\right)
e^{-\mu t} \nonumber \\
&& 
\end{eqnarray}

\subsection*{Identifying $T$, $\lambda t_i $ and $\mu t_i$}
The pattern probabilities obtained above are sufficient for the
estimation of model parameters and the scaled edge lengths.
From Equations~(\ref{eq_a}) and~(\ref{eq_b})
we can write:
\begin{eqnarray*}
\label{model2}
\lambda \beta & = & \frac{p(B)}{p(A)} \\
\left(\frac{\lambda}{\mu}\right) e^{-\mu t}
& = & \frac{p(A)}{1 - \lambda \beta} \,\,=\,\, \frac{p(A)^2}{p(A) - p(B)}
\end{eqnarray*}
Substituting $\lambda \beta $ and 
$ \left(\frac{\lambda}{\mu}\right) e^{-\mu t}$ 
in Equation~(\ref{eq_c_or_d}) we can express
$ \left(\frac{\lambda}{\mu}\right)$ and $e^{-\mu t}$ (and
hence $\mu t$ and $\lambda t$) in terms of $p(A)$,
$p(B)$ and $p(C\vee D)$.

Repeating this exercise for all pairs of leaves
and then applying Lemma~\ref{additive}, we see that
the tree topology and the scaled time parameters
on all edges are a function of the probabilities
of blocks of type $A$, $B$ and $C\vee D$ in all
pairwise alignments. This completes the proof of
Theorem~\ref{invert}.

\section{Invertibility based on sequence lengths}
\label{lengths}
The main result of this section is
\begin{thm}
Given a collection of sequences that have evolved
under a TKF91 process with parameters $\lambda$ and $\mu$
on a phylogenetic tree, the phylogenetic tree and the
scaled edge lengths are uniquely determined by the collection
of sequence lengths provided the sequences are sufficiently long.
\end{thm}

The idea of the proof is that if a sequence of length $X_0$ evolves
under a TKF91 model into a sequence of length $X$ in time $t$,
then assuming $X$ to be the expected length of the sequence
at time $t$ allows us to estimate $\mu t$. To justify this,
we demonstrate  that the relative standard deviation
of the length distribution goes to zero as the initial
length tends to infinity.

Let $P(X,t)$ be the probability that a sequence
evolving under the TKF91 model has length $X$ at
time $t$. The differential equation for $P(X,t)$
is
\begin{equation}
\label{bd}
\frac{dP(X,t)}{dt} = \lambda X P(X-1,t) + \mu (X+1)P(X+1,t)
- \lambda (X+1) P(X,t) - \mu X P(X,t)
\end{equation}
with the initial conditions $P(X = X_0, 0) = 1$ and
$P(X\neq X_0, 0) = 0$.
Observe that this is slightly different from the standard
equation for the simple birth-death process because of
the immortal link, which has 0 death-rate.
Let $M_1(t)$ and $M_2(t)$ denote the first and the second moments of
$P(X,t)$. They are calculated below.

Multiplying both sides of Equation~(\ref{bd}) by $X$, and
summing over $X$, we get
\begin{equation}
\label{m1}
\frac{dM_1}{dt} = (\lambda - \mu)M_1 + \lambda
\end{equation}
with the initial condition
\begin{equation}
M_1(0) = X_0
\end{equation}
This has the solution
\begin{eqnarray}
\label{m1_sol}
M_1(t) &=& \frac{\left((\lambda - \mu)X_0 + \lambda\right)
e^{(\lambda - \mu)t} - \lambda}{\lambda - \mu} \nonumber \\
&=& \frac{\lambda /\mu}{1 - \lambda / \mu}
+ \left(X_0 - \frac{\lambda /\mu}{1 - \lambda / \mu}\right)
e^{-\mu t(1 - \lambda / \mu)}
\end{eqnarray}

Multiplying both sides of Equation~(\ref{bd}) by $X^2$
and summing over $X$, we get the differential equation for
$M_2(t)$.
\begin{equation}
\label{m2}
\frac{dM_2}{dt} = 2(\lambda - \mu)M_2 + (3\lambda + \mu)M_1 + \lambda
\end{equation}
with the initial condition
\begin{equation}
M_2(0) = X_0^2
\end{equation}
This has the solution
\begin{eqnarray}
\label{m2_sol}
M_2(t)& = & \frac{(\lambda + \mu)\lambda}{(\lambda - \mu)^2}
- \frac{\left((\lambda - \mu)X_0 + \lambda\right)(3\lambda + \mu)
}{(\lambda - \mu)^2}e^{(\lambda - \mu)t} \nonumber \\
&+& \left( X_0^2 + 
\frac{\left((\lambda - \mu)X_0 + \lambda\right)(3\lambda + \mu)}
{(\lambda - \mu)^2} - \frac{(\lambda + \mu)\lambda}{(\lambda - \mu)^2}
\right)e^{2(\lambda - \mu)t}
\end{eqnarray}
Therefore, the variance is given by
\begin{eqnarray}
\label{var}
\sigma^2 = M_2 - M_1^2 &=&
\frac{(\lambda + \mu)(\lambda - \mu)X_0 + \lambda^2}
{(\lambda - \mu)^2}e^{2(\lambda - \mu)t}\nonumber \\
&-& \frac{(\lambda + \mu)((\lambda - \mu)X_0 + \lambda)}
{(\lambda - \mu)^2}e^{(\lambda - \mu)t} + 
\frac{\lambda \mu}{(\lambda - \mu)^2}
\end{eqnarray}

The expected length $\bar{L}$ at equilibrium is given by 
$\frac{\lambda / \mu}{1 - \lambda/ \mu}$.
So we calculate $\sigma^2/\bar{L}^2$ as follows.
\begin{eqnarray}
\label{bound}
\frac{\sigma^2}{\bar{L}^2} &=& 
\frac{(\lambda + \mu)(\lambda - \mu)X_0}{\lambda^2}
(e^{2(\lambda - \mu)t} - e^{(\lambda - \mu)t})
+ \left(
(e^{2(\lambda - \mu)t} - \frac{\lambda + \mu}{\lambda}e^{(\lambda - \mu)t}
+\frac{\mu}{\lambda}
\right)\nonumber \\
\end{eqnarray}
This tends to 0 as $\lambda / \mu $ tends to 1.
Therefore, for sufficiently long sequences the tree
and scaled time parameters can be estimated as follows.

First we estimate $\frac{\lambda}{\mu}$.
\begin{equation}
\label{ratio}
\frac{\lambda}{\mu} = \frac{\bar{L}}{\bar{L}+1}
\end{equation}

Now for any two sequences $S_1$ and $S_2$, with lengths
$X_1$ and $X_2$, respectively, $\mu t = \mu (t_1+t_2)$
(which is the scaled time separating $S_1$ and $S_2$)
is estimated from Equation~(\ref{m1_sol}) by substituting
$X_0 = X_1$ and $M_1(t) = X_2$.
Repeating this for all  pairs of sequences, 
and applying Lemma~\ref{additive}, the tree $T$ and
the scaled time parameters $\lambda t_i$ and
$\mu t_i$ are uniquely recovered.

\noindent {\bf Remark.} A weaker form of the results in the
appendix is implicitly assumed here: for every pair of sequences,
we treated one of them to be the ancestor of the other.

\section{Discussion} This paper gives a mathematical analysis of the
TKF91 model on a tree. We present two uniqueness results
for sufficiently long sequences:
the first one establishes a correspondence between
the distribution of certain patterns in a multiple alignment and the
phylogenetic tree and scaled time parameters associated with
edges of the tree; the second result gives a correspondence
between sequence lengths and the phylogenetic tree along with
other model parameters. Results presented here give us
consistency criteria that the alignments and trees constructed
under the TKF91 model from a set of sequences must meet.

It is not likely that the idea of the proofs
presented here will be useful as it is for building trees. The
limiting situation of sequence lengths going to infinity
arises simultaneously with the ratio $\lambda/\mu$ going
to one, and we do not know if the real sequences can be
modelled with the ratio close to one. Nevertheless, the
correspondence between the homology structures and the
trees has useful algorithmic implications. Lunter et al.
\cite{lmdjh_2003} mention that a good method of
sampling multiple alignments under a fixed tree is
currently missing. The correspondence between the homology
structures and the trees could be useful
in an alignment sampling procedure. Many alignments
could be discarded purely on the basis of pattern probabilities
without actually evaluating the alignment likelihoods.
There are two more directions in which future progress may
be made. For substitution models, Erd\"{o}s, et al. \cite{essw_1999}
have studied the problem of finding a lower bound on sequence
lengths that would be sufficient for the correct
reconstruction of the tree with a high probability.
Interestingly this bound is not large: for $n$ taxa,
sequence lengths of the order of a power of $\text{log}\, n$ are
sufficient. It is conceivable that such a result exists
for TKF type processes. Another direction is to prove
similar results for the variants of TKF model, such as the
model proposed by Metzler et al. \cite{mfwh_2005}
in which $\lambda$ and $\mu$ are the same and the variants in
which longer segments are inserted or deleted, (for example,
the TKF92 model and the model of \cite{mlh_2004}).

\section*{Appendix: General Homology Structures}
\label{reversal}

A purpose of this appendix is to demonstrate that
the likelihood of a general homology structure on sequences
related by a TKF91 process on a tree does
not depend on the location of the root on the tree.

Let $[i,j]$ denote the set of integers from $i$ to $j$.
A {\em rooted phylogenetic $X$-tree } $\mathcal{T}$ is a rooted tree with
leaf set $X = \{x_i; \,i \in [1,p]\}$,
root vertex $x_0$ of degree at least 2, 
and all other vertices (denoted by $x_i; i \in [p+1, p+q]$)
of degree at least 3. Let $V$ and $E$ denote, respectively,
the vertex set and the edge set of $\mathcal{T}$. The edges of
$\mathcal{T}$ are directed away from the root.
Associated with each vertex $x_i$ is a sequence $S_i$ of length $L_i+1$
over a finite alphabet $\Sigma $. The sequences are related
to each other by a TKF91 process with parameters $\lambda $, $\mu $
and time parameter $t_e$ for each edge $e$ of $\mathcal{T}$.
The sequences $S_i; i \in [1,p]$ are observed sequences,
and remaining sequences are unobserved.

Let $s_{ij}$ denote the $j$-th character in $S_i$.
A segment of $S_i$ from $s_{ij}$ to $s_{ik}$ is denoted
by $S_i[j,k]$.
The characters $s_{ij}$ and $s_{kl}$ are said to be {\em homologous}
(denoted by $s_{ij}\sim s_{kl}$) if
$x_i$ and $x_k$ have a common ancestor $x_h$ such that a character in
$S_h$ survives as characters $s_{ij}$ and $s_{kl}$ under the TKF91 process.
For notational convenience we think of the immortal link as
the zeroth character in all sequences, and write $s_{i0}\sim s_{j0}$
for all $i$ and $j$. Thus each $S_i$ has $L_i$ characters other than the
immortal link.
By definition, each character in each sequence is homologous to itself.
Therefore, $\sim $ is an equivalence relation on
$\{s_{ij};\, i \in [0,p+q] \}$.
The collection of equivalence classes (or their restriction to
a set of sequences $\{S_i; i \in I \subseteq [0,p+q]\}$)
is called a {\em homology structure} on the sequences \cite{lmdjh_2003},
and is denoted by $\mathcal{H}^{*}$ (or $\mathcal{H}^I)$.

For a character `-' not in $\Sigma $ (called the {\em gap character}),
let $\Sigma^\prime = \Sigma \cup \{-\}$.
An alignment of $S_i;\,i \in I\subseteq [0,p+q]$ is a
collection $\{R_i = (r_{ij}); \,i \in I\}$
of sequences over $\Sigma^\prime $
such that
\begin{enumerate}
\item deleting gap characters from $R_i$ gives $S_i$
\item all $R_i$ have the same length
\item for each $j$, there is an $i$ such that $r_{ij}$ is a non-gap character
\item  for each $j$, non-gap characters $r_{ij}$ constitute an
equivalence class of $\mathcal{H}^I$.
\end{enumerate}
Each alignment corresponds to a unique homology structure,
so each homology structure is represented by a representative
alignment.

For a directed edge $(x_i,x_j)$ of $\mathcal{T}$
let $\mathcal{H}^{ij}$ denote the homology structure
obtained by restricting $\mathcal{H}^{*}$ to the sequences $S_i$
and $S_j$. Let $P(\mathcal{H}^{*})$ denote the probability
of $\mathcal{H}^{*}$.

\begin{prop}
\label{prop-ph*}
\begin{equation}
P(\mathcal{H}^{*}) = P(L_0)\prod_{(x_i,x_j) \in E}P(\mathcal{H}^{ij}|L_i)
\end{equation}
where $P(L_0)$ is the equilibrium probability of $L_0$ given
by
\begin{displaymath}
p(L_0) = \left(1-\frac{\lambda}{\mu}\right)
\left(\frac{\lambda}{\mu}\right)^{L_0}
\end{displaymath}
\end{prop}
\begin{proof}
This follows from the fact that the sequences evolve independently along
the edges of the tree.
\end{proof}

Let $\mathcal{H}^{obs}$ denote a homology structure on the
observed sequences. Its probability $P(\mathcal{H}^{obs})$
is given by the following
\begin{prop}
\label{prop-ph-obs}
\begin{equation}
P(\mathcal{H}^{obs}) = \sum P(\mathcal{H}^{*})
\end{equation}
where the summation is over all homology structures $\mathcal{H}^{*}$
on $S_i;\,i \in [0,p+q]$ whose restriction on $S_i;\,i \in [1,p]$
is $\mathcal{H}^{obs}$.
\end{prop}

\begin{proof}
The R.H.S. is a sum of the probabilities of
all possible histories that result in the homology structure
$\mathcal{H}^{obs}$.
\end{proof}

In this appendix we want to prove that $P(\mathcal{H}^{obs})$
does not depend on the location of $x_0$.
The idea is to apply Felsenstein's {\em pulley principle}
\cite{fel_1981}. But to be able to do that,
we first have to prove the pulley principle when there are
just two observed sequences.

Let $S_i; i \in [0,2]$ be three sequences,
$S_0$ being the ancestor that evolves to
$S_i$ in time $t_i$, for $i \in [1,2]$,
and let $t = t_1 + t_2$.

A homology structure on two sequences can be decomposed into blocks
of characters bounded on the left by homologous characters.
In the case of the leftmost block, the immortal links constitute
the left boundary. This is formally defined below.

We introduce the term {\em related} only for the
analysis of $S_0$, $S_1$ and $S_2$:
two characters $s_{ia}$ and $s_{jb}$ are said to be {\em related}
if they are descendents of the same character in $S_0$.
By definition, each character in $S_0$ is its own descendent
and parent.
A set of segments $\{S_0[a,a+u], S_1[b,b+v], S_2[c,c+w]\}$ is
called a block of $\mathcal{H}^*$ if the following conditions
hold
\begin{enumerate}
\item $s_{0a}\sim s_{1b}\sim s_{2c}$ 
(Note: by convention, $s_{i0}\sim s_{j0}$.)
\item together the three segments are closed under {\em relatedness}.
\end{enumerate}
A two element subset $\{S_i[a,a+u], S_j[b,b+v]\}$ of a block
is called a block of $\mathcal{H}^{ij}$. The event that
the set $\{S_0[a,a+u], S_1[b,b+v], S_2[c,c+w]\}$ is a block
is denoted by $B^*(a,u,b,v,c,w)$.
The event that the set $\{S_i[a,a+u], S_j[b,b+v]\}$ is a block
is denoted by $B^{ij}(a,u,b,v)$.
Probabilities of these events are computed by
summing over the probabilities of all possible contributing
events.

\begin{lem}
\label{lem-pb}
For $i \in [1,2]$,
\begin{eqnarray}
\label{eq-pb}
\lefteqn{
P(B^{0i}(a,m,b,n)| s_{ib}\sim s_{0a},\,L_0 \geq a+m)
}\nonumber \\[5mm]
& = &
\sum_{e = 0}^{m}\ncr{m}{e}(\mu \beta_i)^e\ncr{n}{m-e}
(1-\mu \beta_i)^{m-e}(\lambda \beta_i)^{n-m+e}
(1-\lambda \beta_i)^{m-e+1}
\nonumber \\[-2mm]
\end{eqnarray}
\end{lem}
\begin{proof}
This follows from the solutions of the TKF91 model given in
Lemma~\ref{lem-sol}.
A term in the above summation gives a contribution to the
probability when $e$ characters in $S_0[a,a+m]$ have no descendents in
$S_i$. The remaining $m-e+1$ characters in $S_0[a,a+m]$
have $n+1$ descendents in $S_i$ (counting $s_{ib}$, which is
a descendent of $s_{0a}$). Therefore, there are $m-e+1$ possible
groups of siblings in $S_i$, and $\ncr{n}{m-e}$ ways of making groups.
Each group contributes a factor $(1-\lambda \beta_i)$.
The probability that a character in $S_0$ has
at least one descendent in $S_i$ is $1-\mu \beta_i$.
Therefore, all groups, except the first group that corresponds to the
descendents of $S_{0a}$, contribute a factor $(1-\mu \beta_i)$.
A group of $k>0$ siblings contributes a factor $(\lambda \beta_i)^{k-1}$.
Thus all groups of siblings together contribute the factor
$(\lambda \beta_i)^{n-m+e}$.
Note that there is no change in the formula even when $a=b=0$.
\end{proof}

The above probability does not depend on $a$ and $b$. It is simply
the probability that $n+1$ characters in the ancestral sequence
have $m+1$ descendents after time $t_i$, with the first
character in the ancestor surviving after time $t_i$.
Let this probability be denoted by $P(n\rightarrow m,t_i)$.

\begin{cor}
\label{cor-pb1}
For $i \in \{1,2\}$,
\begin{equation}
\label{eq-pb1}
(\lambda/\mu)^nP(n\rightarrow m,t_i) = 
(\lambda/\mu)^mP(m \rightarrow n,t_i)
\end{equation}
\end{cor}
\begin{proof}
When the probabilities in the above equation
are written using Lemma~\ref{lem-pb}, the coefficients of
$(1-\mu \beta_i)^{k}(1-\lambda \beta_i)^{k}$
on the two sides of the above equation
are equal for each $k$.
\end{proof}

\begin{cor}
\label{cor-pb2}
\begin{eqnarray}
\label{eq-pb2}
&&\sum_{n=0}^{\infty}
(\lambda/\mu)^n P(n\rightarrow m_1, t_1)P(n\rightarrow m_2, t_2)
\nonumber \\[5mm] 
&=& 
\sum_{n=0}^{\infty}
(\lambda/\mu)^{m_1}P(m_1\rightarrow n, t_1)P(n\rightarrow m_2, t_2)
\nonumber \\[5mm] 
&=& (\lambda/\mu)^{m_1}P(m_1\rightarrow m_2, t_1+t_2) \nonumber \\[5mm]
&=&
(\lambda/\mu)^{m_1}
\sum_{e = 0}^{m_1}\ncr{m_1}{e}\ncr{m_2}{m_1-e}(\mu \beta)^e
(1-\mu \beta)^{m_1-e}(\lambda \beta)^{m_2-m_1+e}
(1-\lambda \beta)^{m_1-e+1}\nonumber \\ [5mm]
&=& (\lambda/\mu)^{m_1}P(B^{12}(b,m_1,c,m2)|
s_{1b}\sim s_{2c},\,L_1 \geq b+m_1)
\end{eqnarray}
\end{cor}
\begin{proof}

The second and the third lines follow from Corollary~\ref{cor-pb1}.
The forth line follows from Lemma~\ref{lem-pb}
and the fact that the expression in the third line is
a function of $t_1+t_2$.
If $s_{1b}\sim s_{2c}$ then there is some character
$s_{0a}$ in $S_0$ such that $s_{1b}\sim s_{0a} \sim s_{2c}$.
This implies the last line.
\end{proof}

Let $N^{ij}(a,m,b,n)$ denote the event that
$B^{ij}(a,m,b,n)$ AND $s_{ip}\nsim s_{jq}$
for $a < p < a+m+1$ and $b < q < b+n+1$.
Informally speaking, it is an event that represents a block
with only one pair of homologous characters.

The following lemma relates the probabilities of $B$-type and $N$-type
events.
\begin{lem}
\label{lem-b-n}
\begin{eqnarray}
\label{eq-b-n}
&& P(B^{12}(b_0,m,c_0,n)|s_{1b_0}\sim s_{2c_0},\,L_1 \geq b_0+m)
\nonumber \\[5mm]
&=& 
P(N^{12}(b_0,m,c_0,n)|s_{1b_0}\sim s_{2c_0},\,L_1 \geq b_0+m)
\nonumber \\[5mm]
&+& \sum_{\substack {1 \leq k \leq min(n,m)\\[1mm]
          n_i| \sum_{i=0}^{k} n_i = n-k\\[1mm]
          m_i| \sum_{i=0}^k m_i = m-k}}
X^k
\prod_{i=0}^{k}
P(N^{12}(b_i,m_i,c_i,n_i)|s_{1b_i}\sim s_{2c_i},\,L_1 \geq b_i+m_i)
\nonumber \\[-3mm]
\end{eqnarray}
where 
$c_i = c_{i-1} +  n_{i-1} + 1$ and
$b_i = b_{i-1} +  m_{i-1} + 1$ for $i \geq 1$.
\end{lem}
\begin{proof}
We have considered all possible ways in which a $B(\ldots)$-type
event is partitioned into $N(\ldots)$-type events. The index $k$
denotes the number of pairs of homologous characters in $S_1[b_0,b_0+m]$
and $S_2[c_0,c_0+n]$ other than the homologous pair 
$s_{1b_0}\sim s_{2c_0}$. 
Therefore, each value of $k$ partitions the $B(\ldots)$-type block
in $N(\ldots)$-type blocks. Histories contributing
to each $N(\ldots)$-type block are independent. The term corresponding
to the $k=0$ case is written separately as the first term because
that is what we would like to compute.
\end{proof}
\begin{lem}
\label{lem-pn}
\end{lem}
\begin{eqnarray}
\label{eq-pn}
&& P(N^{12}(b,m,c,n)|
s_{1b}\sim s_{2c},\,L_1 \geq b+m) \nonumber \\[5mm]
&=& 
\sum_{e = 0}^{m}\ncr{m}{e}\ncr{n}{m-e}(\mu \beta)^e
(1-X-\mu \beta)^{m-e}(\lambda \beta)^{n-m+e}
(1-\lambda \beta)^{m-e+1} \nonumber \\
\end{eqnarray}
\begin{proof}
It follows from Corollary~\ref{cor-pb2} that the LHS of
Equation~(\ref{eq-b-n}) is a function of $t_1+t_2$.
We prove by induction on $m+n$ that 
$P(N^{12}(b_0,m,c_0,n)|s_{1b_0}\sim s_{2c_0},\,L_1 \geq b_0+m)$
is a function of $t_1+t_2$. Probabilities $p(A)$, $p(B)$ and
$p(C\vee D)$ computed in Section~\ref{multiple} are functions of
$t_1+t_2$, and correspond to $m+n = 0, 1$ and $2$, respectively,
and are used as the base case of induction.
Let $P(N^{12}(b_0,m,c_0,n)|s_{1b_0}\sim s_{2c_0},\,L_1 \geq b_0+m)$
be a function of $t_1+t_2$ for $m+n < r$, where $r > 2$.
Let $m+n = r$ in Equation~(\ref{eq-b-n}).
Since $m_i+n_i < r$ for all terms in the summation, and
the L.H.S. of Equation~(\ref{eq-b-n}) is a function of $t_1+t_2$,
the only remaining term (the first term on the R.H.S. of 
Equation~(\ref{eq-b-n})) is a function of $t_1+t_2$.

Once the L.H.S. of Equation~(\ref{eq-pn})
is proved to be independent of the root location, the
R.H.S. is written by treating $x_1$ as the ancestor and
arguing in the same manner as in Lemma~\ref{lem-pb}.
\end{proof}

\begin{lem}
\label{lem-2seq-root-ind}
Let $\mathcal{H}^{12}$ be a homology structure on $S_1$ and $S_2$.
Then $p(\mathcal{H}^{12})$ is a function of $t_1+t_2$.
\end{lem}
\begin{proof}
A homology structure has a natural decomposition into
blocks each of which has only one pair of homologous
characters. Events in distinct blocks are independent.
Thus the result follows from Lemma~\ref{lem-pn}

\end{proof}

\begin{thm}
\label{thm-root-ind}
$p(\mathcal{H}^{obs})$ is independent of the root location.
\end{thm}

\begin{proof}
This follows from Propositions~\ref{prop-ph*} and ~\ref{prop-ph-obs},
and Lemma~\ref{lem-2seq-root-ind}.
\end{proof}

\noindent {\bf Remark} The proofs presented here
(in particular the proof that $P(N^{12}(\ldots)$
is a function of $t_1+t_2$) appear somewhat convoluted.
We could have written results similar to
Lemma~\ref{lem-pb} and Corollary~\ref{cor-pb1} for the probabilities
of the $N(\ldots )$ type events. For example, the following corollary
is similar to Corollary~\ref{cor-pb1}.
\begin{cor}
\label{cor-pn1}
For $i \in \{1,2\}$,
\begin{eqnarray}
\label{eq-pn1}
&&(\lambda/\mu)^nP(N^{0i}(a,n,b,m)| s_{ib}\sim s_{0a},\,L_0 \geq a+n) 
\nonumber \\[5mm]
&=& (\lambda/\mu)^mP(N^{0i}(a,m,b,n)| s_{ib}\sim s_{0a},\,L_0 \geq a+m)
\end{eqnarray}
\end{cor}
But this is not useful to claim that $P(N^{12}(\ldots))$ is a
function of $t_1+t_2$. For example, we cannot prove something like
Corollary~\ref{cor-pb2} for $P(N^{12}(\ldots))$
because there may be a situation such as $s_{1i} \nsim s_{2j}$,
$s_{1i}\sim s_{0k}$ and $s_{0k} \nsim s_{2j}$.
As a result the second step of Corollary~\ref{cor-pb2}
(derivation of the 3rd line from the 2nd) fails for $P(N^{12}(\ldots))$.
In fact, Equation~(\ref{eq-pn1}) cannot be regarded as a ``detailed
balance condition'' even if it looks like one.
If a Markov process has transition probabilities given by $P_{ij}$
and if $\pi_i$ is the stationary distribution, then the
process is reversible if and only if $\pi_i P_{ij} = \pi_j P_{ji}$.
But in our problem, $P(N^{12}(\ldots))$
is not a probability of transition from state 1 to state 2. It is
merely a probability of observing certain configuration that is
defined by parameters that take values 1 and 2.
A similar condition has been used as a detailed balance condition
in \cite{mlh_2004} (see Equation (30) on  p539).

Theorem~\ref{thm-root-ind} has been believed to be true by
researchers, and has been used in \cite{lmdjh_2003} to test
correctness of implementation of algorithms. In general such
a result should not be taken for granted as a consequence of
reversibility of the model and the {\em Pulley Principle}
of Felsenstein. In fact, a result such as Theorem~\ref{thm-root-ind}
{\em implies} the pulley principle.

\subsection*{Acknowledgements}
The idea of looking at sequence lengths in
Section~\ref{lengths} was suggested by Mike Steel.
I would like to thank him for many useful suggestions
during the progress of this work.

\end{document}